\documentstyle[12pt,aps,prd]{revtex}

\newcommand{\beq}{\begin{equation}}
\newcommand{\eeq}{\end{equation}}

\def\lap{\lower.5ex\hbox{$\; \buildrel < \over \sim \;$}}
\def\gap{\lower.5ex\hbox{$\; \buildrel > \over \sim \;$}}

\input epsf
\begin{document}

\title{The Cosmological Constant and the Time of Its Dominance}
\author{Jaume Garriga\/$^{1,3}$, Mario Livio\/$^2$ 
and Alexander Vilenkin\/$^3$}
\address{
$^1$ IFAE, Departament de Fisics, Universitat Autonoma de Barcelona,\\
08193 Bellaterra (Barcelona), Spain\\
$^2$ Space Telescope Science Institute, 3700 San Martin Drive,\\
Baltimore, MD 21218, USA\\
$^3$ Institute of Cosmology, Department of Physics and Astronomy,\\
Tufts University, Medford, MA 02155, USA}

\maketitle

\begin{abstract}

We explore a model in which the cosmological constant $\Lambda$ and the 
density contrast at the time of recombination $\sigma_{rec}$ are random
variables, whose range and {\it a priori} probabilities are determined by the
laws of physics.  (Such models arise naturally in the
framework of inflationary cosmology.) 
Based on the assumption that we are typical observers, we show that the
order of magnitude coincidence among the three timescales: the time of
galaxy formation, the time when the cosmological constant starts to dominate
the cosmic energy density, and the present age of the universe, finds
a natural explanation.  We also discuss the probability
distribution for $\sigma_{rec}$, and find that it is peaked near the
observationally suggested values, for a wide class of {\it a priori}
distributions. 

\end{abstract}

\section{INTRODUCTION}

During the past year and a half two groups have presented
(independently) strong evidence that the expansion of the universe is
accelerating rather than decelerating \cite{1}. This surprising result comes
from distance measurements to more than fifty supernovae Type~Ia
(SNe~Ia) in the redshift range $z=0$ to $z=1.2$.  While possible
ambiguities related to evolution, and to the nature of SNe~Ia
progenitors still exist \cite{2}, the data are consistent with the 
cosmological constant (or vacuum energy) contributing to the total 
energy density about 70\% of the critical density 
($\Omega_{\Lambda}\simeq0.7$).

At the same time, other methods, and measurements of the anisotropy of
the cosmic microwave background indicate that matter alone contributes
about $\Omega_M\simeq0.3$, which when combined with the cosmological
constant suggests a flat universe \cite{3}.

These findings raise however an extremely intriguing question. It is
difficult to understand why we happen to be living in the first and only
time in cosmic history in which $\rho_M\sim\rho_{\Lambda}$ (where
$\rho_M$ is the matter density, and $\rho_{\Lambda}$ the vacuum energy
density associated with the cosmological constant). That is, why
\begin{equation}
t_0\sim t_{\Lambda}~~,
\label{1}
\end{equation}
where $t_0$ is the present time and $t_{\Lambda}$ is the time at which
the cosmological constant starts to dominate. Observers living at $t\ll
t_{\Lambda}$ would find $\Omega_M\approx1$ ($\Omega_{\Lambda}\approx0$),
while observers at $t\gg t_{\Lambda}$ would find
$\Omega_{\Lambda}\approx1$
($\Omega_M\approx0$).

There is another, less frequently discussed ``coincidence'', which
also calls for an explanation.  Observationally, the epoch of structure
formation, when giant galaxies were assembled, is at $z\sim 1-3$, or
$t_G\sim t_0/3-t_0/8$.  For the value of $\Lambda$ suggested by
observations, this is within one order of magnitude of $t_\Lambda$,
\begin{equation}
t_G\sim t_\Lambda.
\label{2}
\end{equation}
It is not clear why these seemingly unrelated times should be
comparable.  We could have for example $t_G\ll t_\Lambda$.

In the present work, we explore whether the above ``coincidences''
[eqs.~(\ref{1}) and (\ref{2})]  
could be due to anthropic selection effects.
The approach that we use is one in which it is assumed that some of 
the constants of 
nature are actually random variables, whose range and a~priori 
probabilities are nevertheless determined by the laws of physics. Under 
this assumption, some values which are allowed in principle, may be
incompatible with the very existence of observers. Hence, such values of
the constants cannot be measured.  The values in the observable range
will be measured by civilizations in different parts of the universe,
and we can define the probability $d{\cal P}=P(\chi)d\chi_1 ... d\chi_n$ 
for variables
$\chi_a$ to be in the intervals $d\chi_a$ as being proportional to the
number of civilizations 
that will measure $\chi_a$ in those intervals.  
Following Ref.\cite{AV95}, we shall use the ``principle of mediocrity'',
which assumes that we are ``typical'' observers.  Namely,
we can expect to observe the most probable values of $\chi_a$.

An immediate objection to this approach is that we are ignorant about
the origin of life, let alone intelligence, and therefore the number
of
civilizations cannot be calculated. However, the approach can still be
used to find the probability distribution for parameters which do not
affect
the physical processes involved in the evolution of life. The
cosmological constant $\Lambda$ and the
amplitude of density fluctuations at horizon crossing 
$Q$ are examples of such
parameters. If the parameters $\chi_a$ belong to this
category, then the
probability for a carbon-based 
civilization to evolve on a suitable planet is 
independent of $\chi_a$, and instead of the number of civilizations we
can use
the number of habitable planets or, as a rough approximation, the
number of suitable galaxies.  
We can then write
\begin{equation}
P(\chi)d^n\chi\propto d{\cal N},
\label{3}
\end{equation}
where $d{\cal N}$ is the number of galaxies that are 
formed in regions where $\chi_a$ take values in the intervals
$d\chi_a$.

The problem of calculating the probability distribution $d{\cal
P}(\chi)$ can be split into two parts.  The number of galaxies
$d{\cal N}(\chi)$ in Eq.~(\ref{3})
is proportional to the volume of the comoving
regions where $\chi_a$ take specified values and to the density of
galaxies in those regions.  The volumes and the densities can be
evaluated at any time.  Their product should be independent of the
choice of this reference time, as long as we include both galaxies
that formed in the past and those that are going to be formed in
the future.  For some purposes
it is convenient to evaluate the volumes and the densities
at the time of recombination, $t_{rec}$.  We can then write
\begin{equation}
d{\cal P}(\chi)= \nu(\chi)d{\cal P}_*(\chi) .
\label{4}
\end{equation}
Here, $d{\cal P}_*(\chi)=P_*(\chi)d^n\chi$ 
is proportional to the volume of those
parts of the universe where $\chi_a$ take values in the intervals
$d\chi_a$, and
$\nu(\chi)$ is the average number of galaxies 
that form per unit 
volume with cosmological parameters specified by the values of
$\chi_a$.  $d{\cal P}_*(\chi)$ is an {\it a priori} probability
distribution\footnote{We use the term {\it a priori} in the sense that
this distribution is independent of the existence of observers.} 
which should be determined from the theory of initial
conditions (e.g., from an inflationary model).  On the other hand,
the calculation of $\nu(\chi)$ is a standard astrophysical problem,
unrelated to the calculation of the volume factor $d{\cal P}_*(\chi)$.

The principle of mediocrity (which is closely related to the
``Copernican principle'') has been applied to determine the likely
values of the cosmological constant
\cite{AV95,Weinberg,Martel,Efstathiou}, of the density parameter
$\Omega$ \cite{VW,GTV}, and of the density fluctuations at horizon crossing 
$Q$ \cite{Tegmark}.  A very similar approach was
used by Carter \cite{Carter}, Leslie \cite{Leslie} and Gott
\cite{Gott} 
to estimate the expected
lifetime of our civilization.  Gott also applied it 
to estimate the lifetimes of various
political and economic structures, including the journal ``{\it Nature}''
where his article was published.  Related ideas have also been
discussed by Linde {\it et. al.} \cite{Linde} and by Albrecht
\cite{Albrecht}. 

Spatial variation of the ``constants'' 
can naturally arise in the framework of inflationary
cosmology \cite{Lindebook}.  The dynamics of light scalar fields
during inflation are strongly influenced by quantum fluctuations, causing
different regions of the universe to thermalize with different values of
the fields. For example, what we
perceive as a cosmological constant could be a potential $U(\phi)$ of
some field $\phi(x)$.  If this potential is very flat, so that the
evolution of $\phi$ is much slower than the Hubble expansion, then
observations will not distinguish between $U(\phi)$ and a true
cosmological constant.  Observers in different parts of the universe
would then measure different values of $U(\phi)$.
Quite similarly, the potential of the inflaton field $\Phi$ that drives
inflation can depend on a slowly-varying field $\phi$.  In this case,
regions of the universe thermalizing with different values of $\phi$
will be characterized by different amplitudes of the cosmological
density fluctuations.  Examples of models of this sort have been given
in Refs.\cite{GTV,AV98}.

The application of the principle of mediocrity in our case will require
comparing the expected numbers of civilizations in parts of the universe
with different values of $\Lambda$ and $Q$, which will be treated as random
variables. In fact, for our purposes, it will be convenient to deal
with an additional random variable, $t_G$. This is 
because one of the questions we are 
addressing is the coincidence (2), and galaxy formation can itself be
modeled as a random process which takes place over a range of times
for given $Q$ and $\Lambda$.  Instead of $Q$, it will be more
convenient to use the density contrast on the galactic scale at the
time of recombination, $\sigma_{rec}$.
Throughout the paper we assume that the universe is flat,
$\Omega_\Lambda +\Omega_M=1$.

The paper is organized as follows.  We shall first consider the
situation in which only the cosmological constant is allowed to vary, with
all other parameters being fixed.  In Section II we will show that
the most likely values of $\Lambda$ and $t_G$ in this case are such that
$t_\Lambda \sim t_G$.  In Section III we shall argue that the most
likely epoch for the existence of intelligent observers is $t_0\sim t_G$.  
This completes the argument that coincidences (\ref{1}) and (\ref{2})
are indeed  to be expected in this class of models.
In Section IV we discuss models where both $\Lambda$ and $\sigma_{rec}$ 
are
variable and outline the calculation of the probability distribution
for $t_\Lambda$ and $t_G$.  In our analysis of these models we go
beyond the issue of the cosmic time coincidence and discuss the
values of $t_\Lambda$ and of 
the density contrast $\sigma_{rec}$ detected by typical
observers.  Our conclusions are summarized in Section V.

\section{Why is \lowercase{$t_\Lambda$}\ $\sim$\ \lowercase{$t$}${}_G$?}

In this and the following section we assume that the cosmological
constant $\Lambda$ is the only variable parameter.  
Weinberg \cite{W87} was the first to point out that not all values of
$\Lambda$ are consistent with the existence of conscious observers.
In a spatially flat universe with a cosmological constant, 
gravitational clustering effectively stops at a redshift
$(1+z_\Lambda) \sim (\rho_\Lambda/\rho_{M0})^{1/3}$, when
$\rho_\Lambda$ becomes comparable to the matter density $\rho_M$.
(Here, $\rho_{M0}$ is the present matter density.)  At later times,
the vacuum energy dominates and the universe enters a de Sitter stage
of exponential expansion.  An anthropic bound on $\rho_\Lambda$ can be
obtained by requiring that it does not dominate before the redshift
$z_{max}$ when the earliest galaxies are formed,
\begin{equation}
\rho_\Lambda\lesssim (1+z_{max})^3\rho_{M0}.
\label{5}
\end{equation}
Weinberg took $z_{max}\sim 4$, which gives $\rho_\Lambda\lesssim
100\rho_{M0}$. 

One expects that the {\it a priori} probability distribution ${\cal
P}_*(\rho_\Lambda)$ should vary on some characteristic particle physics
scale, $\Delta\rho_\Lambda\sim\eta^4$.  The energy scale $\eta$ could
be the Planck scale $\eta_{pl}\sim 10^{19}$ GeV, the grand
unification scale $\eta_{GUT}\sim 10^{16}$ GeV, or the electroweak
scale $\eta_{EW}\sim 10^2$ GeV.  For any reasonable choices of $\eta$
and $z_{max}$, $\Delta\rho_\Lambda$ exceeds the anthropically allowed
range of $\rho_\Lambda$ by many orders of magnitude.  We can therefore
set 
\beq
{\cal P}_*(\rho_\Lambda)=const
\label{flat}
\eeq 
in the range of interest
\cite{W87}.  With this flat distribution, a value of
$\rho_\Lambda$ picked randomly from an interval
$|\rho_\Lambda|\lesssim \rho_\Lambda^m$ is likely to be comparable to
$\rho_\Lambda^m$ (the probability of picking a much smaller value is
small).  In this sense, the flat distribution (\ref{flat})
favors larger values of $\rho_\Lambda$.

The anthropic bound (\ref{5}) specifies the value of $\rho_\Lambda$
which makes galaxy formation barely possible.  However, the principle
of mediocrity suggests that we are most likely to observe not these
marginal values, but rather the ones that maximize the number of galaxies.
This suggests that $\Lambda$-domination should not occur before a
substantial fraction of matter has collapsed into galaxies.  The
largest values of $\Lambda$ consistent with this requirement are such
that $t_\Lambda\sim t_G$.  Hence, the coincidence (\ref{2}) is to be
expected if we are typical observers \cite{Foot1}.

Let us now try to make this more quantitative.  
It will be convenient to introduce a variable 
\begin{equation}
x={\Omega_\Lambda\over{\Omega_M}}=\sinh^2\left({t\over t_{\Lambda}}\right),
\label{x}
\end{equation}
where for convenience, we have defined $t_{\Lambda}$ as the time at
which $\Omega_{\Lambda}= \sinh^2(1)\Omega_M \approx 1.38 \Omega_M$.
At the time of recombination, for values of $\rho_\Lambda$ within the
anthropic range, $x_{rec}\approx \rho_\Lambda/\rho_{rec}\ll 1$, where the
matter density at recombination, $\rho_{rec}$, is independent of
$\Lambda$.  We can therefore express the probability distribution for
$\rho_\Lambda$ as a distribution for $x_{rec}$,
\beq
d{\cal P}(x_{rec})\propto \nu(x_{rec})dx_{rec},
\label{Pxrec}
\eeq
where $\nu(x_{rec})$ is the number of galaxies formed per unit volume
in regions with a given value of $x_{rec}$.   The calculation of the
distribution (\ref{Pxrec}) was discussed in detail by Martel
et. al. \cite{Martel}.  A simplified version of their analysis is
given in the Appendix.

Galaxies form at the time when the density contrast 
(evolved according to the linear theory) exceeds a
certain critical value $\Delta_c(x)$. For small values of $x$, when the
cosmological constant is negligible, we have $\Delta_c(x)\approx 1.69$ as
in the Einstein de Sitter model. However, it is known that 
$\Delta_c$ is slightly dependent
on $x$, with $\Delta_c(\infty)\approx 1.63$. Thus, $\Delta_c$ varies by no 
more than $4\%$ in the whole relevant range, and in what follows we shall 
ignore its $x$-dependence. The number of galaxies wich have assembled up 
to a given time $t$ for a given value of the cosmological constant
(that is, up to a given $x$
for a given value of $x_{rec}$) can thus be estimated as \cite{PS}
\begin{equation}
\nu(x,x_{rec})=
{\rm erfc}\left({\Delta_c\over \sqrt{2}\sigma_{rec}G(x,x_{rec})}
\right).
\label{ps}
\end{equation}
The factor $G(x,x_{rec})=x_{rec}^{-1/3}F(x)$, where 
\beq
F={5\over 6} \left({1+x \over x}\right)^{1/2} 
\int_0^x {d\omega \over \omega^{1/6}(1+\omega)^{3/2}},
\eeq
accounts for the growth of the dispersion in the density contrast 
$\sigma$ on the galactic scale 
from its value $\sigma_{rec}$ at the time of recombination until time
$t(x)$.  For
small $x$ we have $F\approx x^{1/3}$, and perturbations grow as in the
Einstein de Sitter model. However, at large $x$ the growth of perturbations
is stalled and we have $F(\infty)=(5/6)\beta(2/3,5/6)\approx 
1.44$. 
The number of galaxies that will assemble in a given interval
of $x$ will thus be given by
\begin{equation}
d\nu(x,x_{rec}) \propto \exp\left[-{1\over 2}
\left({\Delta_c \over {F(x)}}{x^{1/3}_{rec}\over\sigma_{rec}}\right)^2\right]
{{F'(x)}\over {F^2(x)}} {x^{1/3}_{rec}\over \sigma_{rec}}dx.
\label{distg}
\end{equation}
Multiplying by a flat {\em a priori} distribution for $x_{rec}$, we have
\begin{equation}
d{\cal P}(x,x_{rec})\propto d\nu(x,x_{rec}) dx_{rec}.
\label{prob2}
\end{equation}
The probability for an observer to live in a galaxy that formed in a
given logarithmic interval of $t_G/t_\Lambda$ can now be obtained by
integrating (\ref{prob2}) 
with respect to $x_{rec}$ while 
keeping $x$ fixed. The result is
\beq
d{\cal P}(t_G/t_{\Lambda}) 
\propto \sigma_{rec}^3 F^2 F' {dx\over d\ln(t_G/t_{\Lambda})}
d\ln(t_G/t_\Lambda).
\label{prob3}
\eeq
This distribution is shown in Fig.~1 (curve $a$).  It has a broad peak which
almost vanishes outside of the range
$.1\lesssim (t_G/t_{\Lambda}) \lesssim 10$.  The maximum of the
distribution is at $t_G/t_{\Lambda}\approx 1.7$ and the median value is
at $t_G/t_\Lambda\approx 1.5$. 
Thus, most observers will find that their galaxies
formed at $t\sim t_\Lambda$, and therefore the coincidence
\beq
t_G\sim t_\Lambda
\label{tgtlambda}
\eeq
is explained.

It is also of some interest to consider the distribution (\ref{prob2}) 
without performing any integrations. By changing from the variables $x$ and 
$x_{rec}$ to the variables $t_G$ and $t_\Lambda$ we have
\begin{equation}
d{\cal P}\propto
\sigma_{rec}^3
\exp\left[-{(t_\sigma/ t_\Lambda)^{4/3}\over 2 F^2}\right]
{{F'(x)}\over {F^2(x)}}\left({t_\sigma\over t_\Lambda}\right)^{8/3}
\left({t_G\over t_\Lambda}\right)\sinh\left({2t_G\over t_\Lambda}\right)
d\ln t_G\ d\ln t_\Lambda,
\label{long}
\end{equation}
where $x=x(t_G/t_\Lambda)$ and 
\beq
t_{\sigma}\equiv (\Delta_c/\sigma_{rec})^{3/2} t_{rec} 
\label{tsigma}
\eeq
is the
time at which the density contrast on galactic scales would reach the
critical value $\Delta_c$ in an Einstein-de Sitter model. Here we are
not allowing for variations of $\sigma_{rec}$, and therefore this time
is just a constant. The probability density (\ref{long}) per unit area
in the $(\log t_G,\log t_\Lambda)$ plane is plotted in Fig. 2. Note that
the peak is in the region where $t_G\sim t_\Lambda \sim t_{\sigma}$.
Different projections of this plot are useful. If we integrate
along the vertical axis, then we obtain the probability distribution
for the time when $\Lambda$ dominates, which is equivalent to 
(\ref{Pxrec}), whereas if we integrate diagonally along 
$(t_G/t_{\Lambda})=const$
lines, we obtain (\ref{prob3}).

\begin{figure}[t]
\centering
\hspace*{-4mm}
\leavevmode\epsfysize=10 cm \epsfbox{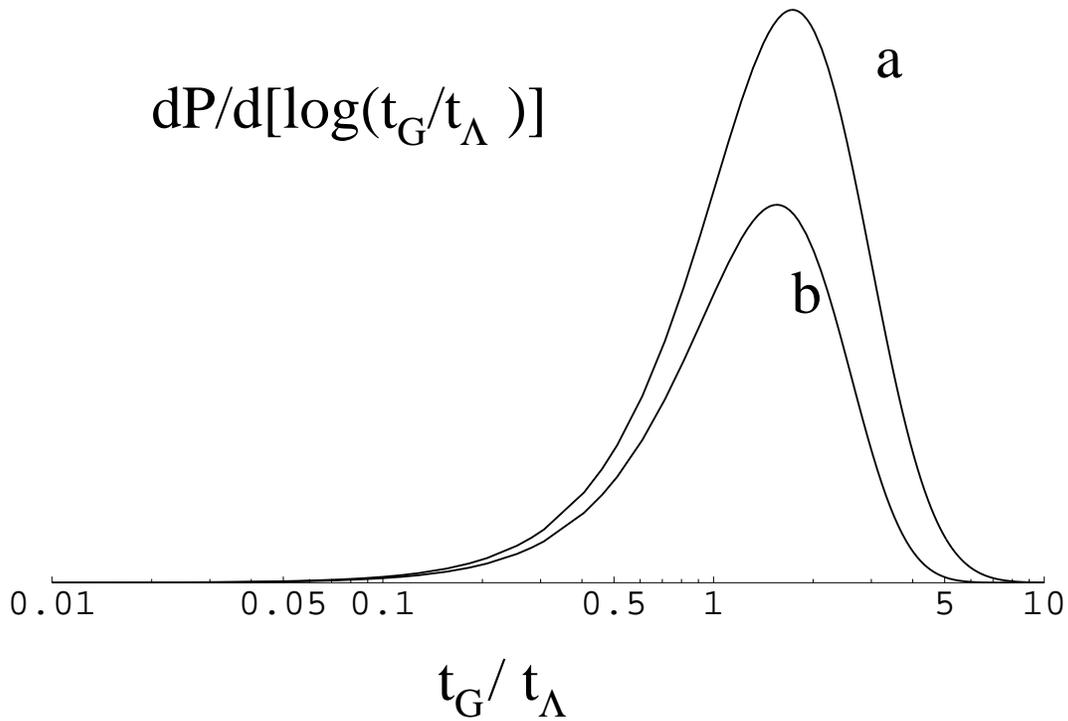}\\[3mm]
\caption[fig1]{\label{fig1}The probability density per unit 
logarithmic interval of
$t_G/t_{\Lambda}$, Eq.(\ref{prob3}), is shown (curve $a$). The 
maximum is at $t_G/t_{\Lambda}\approx 1.7$ whereas the median value is
at $t_G/t_\Lambda\approx 1.5$. The same distribution taking into account the
cooling boundary $t_{cb}$ discussed in Section IV is also plotted 
(curve $b$). The parameters have been chosen so that $t_{cb}=.5\ t_\sigma$
[see Eq. (\ref{nu'})]. }
\end{figure}

\begin{figure}[t]
\centering
\hspace*{-4mm}
\leavevmode\epsfysize=10 cm \epsfbox{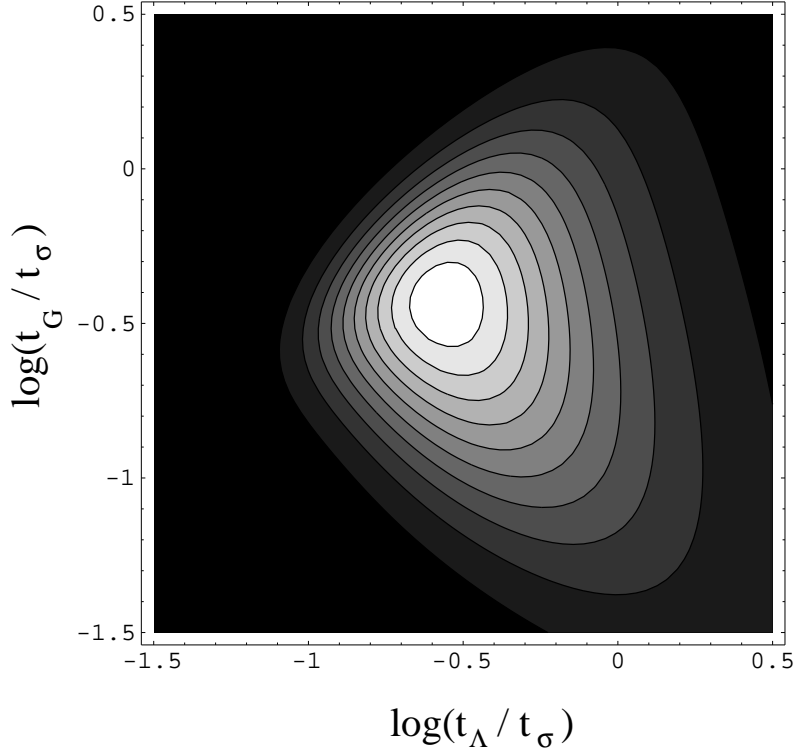}\\[3mm]
\caption[fig2]{\label{fig2} The joint probability density (\ref{long}) per unit
area in the plane $\log\ (t_\Lambda/t_\sigma)$ (horizontal axis) 
$\log (t_G/t_\sigma)$ (vertical axis), where $t_{\sigma}$ is defined in 
Eq.~(\ref{tsigma}).  }
\end{figure}

\section{Why Now?}

As we noted in the Introduction, one of the most puzzling aspects of the 
value of $\Omega_{\Lambda}$ is related to the fact that the coincidence
$t_0\sim t_{\Lambda}$ appears to be implying that we live in a special
time. A similar problem exists even if a quintessence component
\cite{Quintessence} is assumed (see Section V). 
As we have shown in Section II, the epoch when giant galaxies
are assembled, $t_G$, is expected to roughly coincide with the epoch of
cosmological constant dominance, $t_{\Lambda}$. Therefore, if we could
explain why $t_0\sim t_G$, the puzzle of the cosmic age coincidence
would be resolved.

Most of carbon-based life may be expected
to have appeared in the universe around the peak in the universal carbon
production rate, at $t_{\rm carbon}$. The main contributors to carbon in
the interstellar medium are stars in the mass range 1--2~M$_{\odot}$,
through carbon stars and planetary nebulae \cite{XX}. Consequently, detailed
simulations \cite{XXX} show that the peak in the cosmic carbon production
rate 
is delayed only by less than a billion years compared to the peak in the 
cosmic star formation rate, $t_{\rm SFR}$, namely, 
\begin{equation}
t_{\rm carbon}\sim t_{\rm SFR}~~.
\end{equation}

The appearance of intelligent life is further delayed by no more than
a fraction of the main sequence lifetime of stars in the spectral range
mid-F to mid-K (~5-20 Gyr; \cite{XXX,Foot2}). 
Following the main sequence, the expansion
and increase in luminosity of stars spells the end of the possible existence
of a biosphere on planets. Only stars in the above spectral range are
expected to have continuously habitable zones around them (namely, ensuring
the presence of liquid water and the absence of catastrophic cooling by
${\rm CO_2}$ clouds on planetary surfaces; \cite{YY}).  Thus we have
\begin{equation}
t_{\rm IL}\sim t_{\rm carbon}\sim t_{\rm SFR}~~.
\label{til}
\end{equation}

The ``present time'' $t_0$ can be defined as the time when a
civilization evolves to the point where it is capable of measuring the
cosmological constant and becomes aware of the coincidence (\ref{1}).
The experience of our own civilization suggests that, on a
cosmological timescale, this time is not much different from $t_{\rm
IL}$,
\beq
t_0\approx t_{\rm IL}.
\label{t0til}
\eeq

Carter \cite{Carter} and others \cite{Leslie,Gott} used the principle
of mediocrity to argue that the lifetime of our civilization
is unlikely to be much longer than the time it has already existed,
that is, $\sim 10^4$ yrs.  If we are typical, then this should be the
characteristic lifetime of a civilization.  This would imply that
Eq.~(\ref{t0til}) is valid even if $t_0$ is understood as the time
when any astronomical observations can be made.  Carter's argument has
some force, but it is based on a single data point, and one may be
reluctant to accept it, considering in particular its pessimistic
implications.  We note, however, that with our definition of $t_0$,
Eq.~(\ref{t0til}) is likely to be valid regardless of the validity of
Carter's argument (that is, even if civilizations are likely to
survive much longer than $t_{\rm IL}$).  Combining (\ref{t0til}) with
(\ref{til}), we have that for a typical civilization
\beq
t_0\sim t_{\rm SFR}.
\label{t0tsfr}
\eeq

Finally, models of galaxy formation in hierarchical clustering theories
propose that Lyman-break galaxies (at $z\sim3$) are the first objects of
galactic size which experience vigorous star formation \cite{ZZZ}. These 
objects therefore signal the onset of the epoch of galaxy formation, 
with cosmic star formation and galaxy formation being closely linked. 
In fact, the mergers and collisions of ``sub-galactic'' objects to 
produce galactic-size structures, are responsible for the enhanced star 
formation. In hierarchical models therefore
\begin{equation}
t_G\sim t_{\rm SFR}~~.
\label{tgtsfr}
\end{equation}

The above relation is also supported by observations of the star
formation history, showing 
that the star formation rate rises from the present to about $z\sim1$, 
with a broad peak (of roughly constant star formation rate) in the 
redshift range $z\sim1$--3 \cite{ZZ}. This corresponds roughly to 
$t_{\rm SFR}\sim t_0/3$--$t_0/8$, in agreement with Eq.~(\ref{tgtsfr}).
In fact, more than 80\% of the stars have already formed
($\Omega_{gas}/\Omega_{stars}\sim 0.18$ \cite{Fukugita}).

Combining Eqs.~(\ref{tgtlambda}),(\ref{t0tsfr}),(\ref{tgtsfr}) 
above we obtain the desired relation
\begin{equation}
t_0\sim t_G\sim t_{\Lambda}~~.
\end{equation}

\section{Models with variable $\Lambda$ and $\sigma_{rec}$}

In the previous discussion we have assumed a fixed value of
the density contrast at recombination $\sigma_{rec}$ (or equivalently,
a fixed value of $Q$).  
This determines the parameter $t_{\sigma}\equiv 
(\Delta_c/\sigma_{rec})^{3/2} t_{rec}$ appearing in the distribution
(\ref{long}) and therefore, as it is clear from Fig. 2, the most probable
time at which the cosmological constant will dominate 
$t_{\Lambda}\sim t_{\sigma}$ \cite{Martel}. 

If $\sigma_{rec}$ is itself treated as a random variable, with 
a priori distribution ${\cal P}_*(\sigma_{rec})d\ln\sigma_{rec}$ 
then the most probable value of $t_{\Lambda}$ will of course have some 
dependence on ${\cal P}_*$. However, as we shall argue, this
dependence is not too strong provided that ${\cal P}_*$ satisfies
some qualitative requirements, in which case the most probable values
of $t_{\Lambda}$ and $\sigma_{rec}$ are  actually determined by the 
fundamental constants involved in the cooling processes which take
place in collapsing gas clouds. 

\subsection{The cooling boundary}

So far, we have assumed that all the galactic-size objects collapsing
at any time form luminous galaxies. However,
galaxies forming at later times will have a lower density
and shallower potential wells. They are thus vulnerable to losing all their
gas due to supernova explosions \cite{Tegmark}. Moreover, a collapsing cloud
fragments into stars only if the cooling timescale of the cloud
$\tau_{cool}$ is smaller than the collapse timescale
$\tau_{grav}$. Otherwise,
the cloud stabilizes into a pressure supported configuration
\cite{reos,Tegmark}.  The cooling rate of such 
pressure supported clouds is exceedingly low, and it is possible that
star formation in the relevant mass range will be suppressed in these
clouds even when they eventually
cool.  Hence, it is conceivable that galaxies that fail to cool during
the initial collapse give a negligible contribution to $\nu$.
Fragmentation of a cloud into stars will be suppressed after a certain 
critical time which we shall refer to as the ``cooling boundary''
$t_{cb}$ \cite{Tegmark}.  

To determine $t_{cb}$, let us first consider the case of
a matter dominated universe (not necessarily flat) without a 
cosmological constant. An overdensity which is destined to collapse can 
be described in the spherical model as a part of a closed 
Friedmann-Robertson-Walker (FRW) universe. The size of this 
spherical region at the time of recombination is such that it
basically contains the mass of a galaxy. The virialization temperature 
and the density after virialization will be quite independent of what 
happens outside the region, depending only on its gravitational 
energy at the time $t_{vir}$ when it collapses.  The virial velocity 
will then be given by $v_{vir}\sim (GM_{g}/L)^{1/2}$, where $L$ is the 
size of the collapsing object at $t_{vir}$.
The density of the virialized collapsing cloud $\rho_{vir}$ is
given by \cite{Tegmark,laco93}
\beq
\rho_{vir} \sim 10^2 (G t_{vir}^2)^{-1}.
\eeq
The virialization temperature can be
estimated as 
$T_{vir} \sim m_p v_{vir}^2 \sim m_p (G^3 \rho_{vir} M_{g}^2)^{1/3}$.
Here $m_p$ is the proton mass.
The later an object collapses, the colder and more dilute it would be.

If there is a cosmological constant, then these estimates still
hold to good approximation. Indeed, a spherical region will only
collapse if its intrinsic ``curvature'' term is always dominant with
respect to the cosmological constant term. The ``potential'' energy at 
the time of collapse and the properties of the virialized cloud will 
basically remain unaltered.
In principle, a spherical region with a cosmological constant
could enter a ``quasistatic'' phase where the gravitational pull is 
nearly balanced by the repulsion due to the cosmological constant. 
After a long period of time, this region might finally collapse and 
virialize to a large enough temperature. However, since the quasistatic
phase is unstable we shall disregard this marginal possibility.

The cooling rate $\tau_{cool}^{-1}$ of a gas cloud of fixed mass 
depends only on its density and temperature, 
but as shown above both of these quantities are determined by 
$t_{vir}$ \cite{Foot4}.
The timescale needed for gravitational collapse is $\tau_{grav}\sim t_{vir}$. 
Therefore, the condition $\tau_{cool}<\tau_{grav}$ gives an upper 
bound $t_{cb}$ on the time at which collapse occurs. 
Various cooling processes
such as Bremsstrahlung and line cooling in neutral hydrogen and helium
were considered in Ref. \cite{Tegmark}. 
For a cloud of mass $M_g \approx 10^{12} M_{\odot}$, 
cooling turns out to be  efficient \cite{Foot5} for 
\begin{equation}
t<t_{cb} \approx 3 \cdot 10^{10} {\rm yr}.
\label{rhoc}
\end{equation}
In any case, this value of $t_{cb}$ should be taken only as indicative, 
since the present status of the theory of star formation 
does not allow for very precise estimates.

\subsection{Likely values of $t_\Lambda$}

Let us now consider the probability distribution for the 
three independent variables $x, x_{rec}$ and $\sigma_{rec}$. This will be 
proportional to
the number of galaxies forming at a time charachterized by $x$ in a 
region with given values of $\sigma_{rec}$ and $x_{rec}$, 
\begin{equation}
d{\cal P}(x,x_{rec},\sigma_{rec}) \propto 
{\cal P}_*(\sigma_{rec}) \exp\left[-{1\over 2}
\left({\Delta_c \over F}{x^{1/3}_{rec}\over\sigma_{rec}}\right)^2\right]
{F'\over F^2} {x^{1/3}_{rec}\over \sigma_{rec}} dx\ dx_{rec}\
d\ln\sigma_{rec}.  
\label{large}
\end{equation}
Let us assume for simplicity a power-law a priori distribution,
\beq
 {\cal P}_*(\sigma_{rec})\propto 
\sigma_{rec}^{-\alpha}, 
\eeq
where $\alpha$ is a constant. Then we can 
immediately integrate over $\sigma_{rec}$ and obtain
\begin{equation}
d{\cal P}(x,x_{rec}) \propto  x_{rec}^{-\alpha/3}
F^{\alpha-1} F'\ dx\ dx_{rec}.
\end{equation}
Now we can integrate with respect to the ``time'' $x$ at which galaxies
assemble, from the time of recombination $x_{rec}$ to the cooling boundary
\beq
x_{cb}= \sinh^2(t_{cb}/t_{\Lambda}).
\eeq
The integral is simply the difference in $F^{\alpha}$ between the 
two boundaries in the integration range, and we shall neglect
the contribution at $x_{rec}$.
Finally, using $t_{\Lambda}=t_{rec} x_{rec}^{-1/2}$ we 
obtain a probability distribution for $t_{\Lambda}$
\begin{equation}
d{\cal P}(t_{\Lambda}) \propto 
F^{\alpha}(\sinh^2(t_{cb}/t_{\Lambda})) 
t_{\Lambda}^{{2\alpha\over 3}-2} d\ln(t_{\Lambda}/t_{cb}).
\label{tlambda}
\end{equation}
Thus, the most probable value of $t_{\Lambda}$ is determined by
$t_{cb}$ and $\alpha$. 

In Fig. 3, this distribution is plotted for different 
values of $\alpha$ ranging from 4 to 15. In all these cases we have
\begin{equation}
t_{\Lambda}\sim t_{cb}. \label{tltcb}
\end{equation}
The behaviour of the distribution is different 
for $\alpha\leq 3$. Note that $F(y)\propto y^{1/3}$ 
for small y, whereas $F$ saturates at a constant value for
large $y$. This means that if $\alpha<3$, the distribution (\ref{tlambda}) 
would favour very small values of $t_{\Lambda}$. 
The reason is that for a small $\alpha$ 
the {\em a priori} distribution is not too suppressed at large
$\sigma_{rec}$, and
it pays off to increase $\sigma_{rec}$ in order to obtain a large 
number of collapsed objects very soon after recombination. Therefore
the time of $\Lambda$ domination can be very short without
interfering with galaxy formation. Of course, this would 
result in an overwhelming majority of the galaxies in regions which do not 
look anything like ours.
On the other hand, if $\alpha>3$, small values of $\sigma_{rec}$ are 
preferred. However, the value of $\sigma_{rec}$ should at least be
large enough for galaxy formation to occur marginally before the cooling 
boundary $t_{cb}$. Therefore, 
if the cosmological constant is not to interfere with 
galaxy formation, the result (\ref{tltcb}) is expected.  More
generally, we expect the relation (\ref{tltcb}) to be valid if the
{\it a priori} distribution decreases faster than $\sigma_{rec}^{-3}$
at small $\sigma_{rec}$.  With $t_{cb}$
from Eq.~(\ref{rhoc}) and the $t_\Lambda$ suggested by observations, the
relation (\ref{tltcb}) is indeed satisfied.

\begin{figure}[t]
\centering
\hspace*{-4mm}
\leavevmode\epsfysize=10 cm \epsfbox{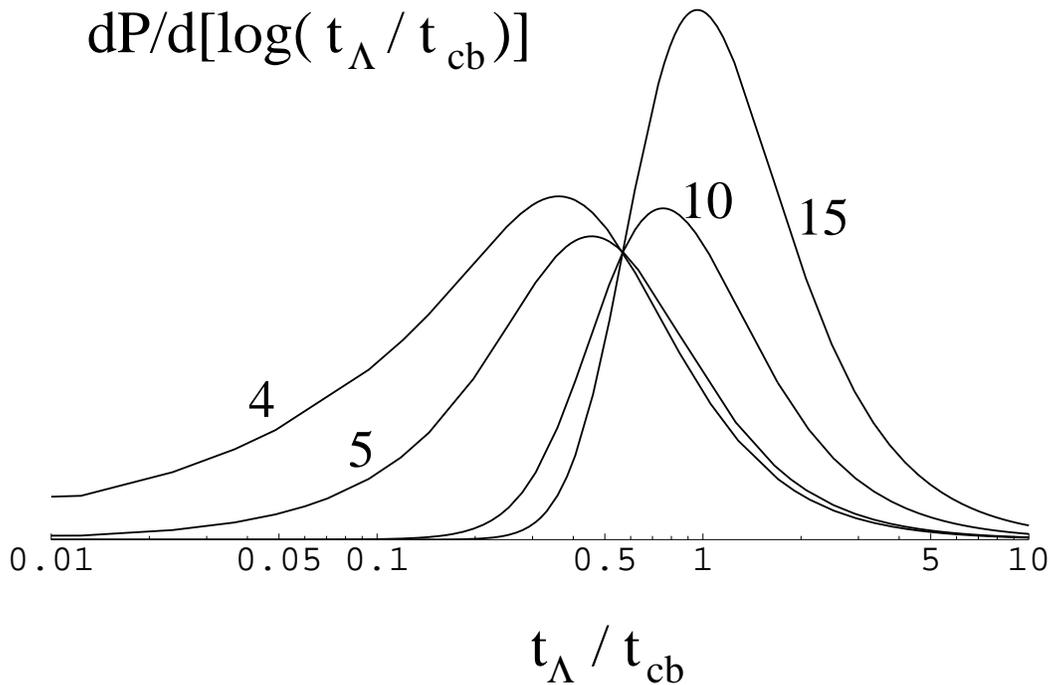}\\[3mm]
\caption[fig3]{\label{fig3} 
 The probability density per unit logarithmic
interval of $t_{\Lambda}/t_{cb}$, Eq. (\ref{tlambda}), for 
different values of the parameter $\alpha$ ($\alpha=4,5,10,$ and $15$).
 }
\end{figure}

\subsection{Likely values of $\sigma_{rec}$} 

A probability distribution for $\sigma_{rec}$ can be obtained by
integrating (\ref{large}) first over $x_{rec}$ over the relevant range
$\sinh(x_{rec}^{1/2} t_{cb}/t_{rec})>x^{1/2}$, and then over $x$. 
The result can be expressed as 
\begin{equation}
d{\cal P}(\beta) \propto \beta^{2(\alpha-3)/3} G(\beta)\
d\ln\beta,
\label{short}
\end{equation}
where we have introduced the $\sigma_{rec}$ dependent parameter
\begin{equation}
\beta={t_\sigma \over t_{cb}}=  
\left({\Delta_c\over\sigma_{rec}}\right)^{3/2} {t_{rec}\over t_{cb}}
\label{nu'}
\end{equation}
and the function
\begin{equation}
G(\beta) = \int_0^{\infty}
\exp\left[-{1 \over 2 F^2}\left({\beta t\over
t_\Lambda}\right)^{4/3}
\right] F' \left[F^2+{1\over 2}
\left({\beta t\over t_\Lambda}\right)^{4/3}\right] dx.
\label{gdistr}
\end{equation}
The function $G(\beta)$ is plotted in Fig. 4 (thick solid line). 
It stays constant for $\beta<1$ (towards large $\sigma_{rec}$) 
and it drops to zero around $\beta \approx 10$. For larger $\beta$ it 
falls off as $\beta^{4/3}\exp(-\beta^{4/3}/2)$. This function is
multiplied in (\ref{short}) by the factor 
$\sigma_{rec}^{3-\alpha}$, which depends on the a priori
distribution for $\sigma_{rec}$. If this factor is a
decreasing function of $\sigma_{rec}$ (i.e. $\alpha>3$) then (\ref{short})
peaks between $1<\beta\lesssim 10$. This is illustrated
in Fig. 4 (thin curves) for $\alpha=4$ and $5$.  From the definition
of $\beta$ we have
\beq
\sigma_{rec}={\Delta_c\over (1+z_{rec})}
\left({2\over 3\beta H_0t_{cb}\sqrt{\Omega_M}}\right)^{2/3}\approx
1.1\cdot 10^{-3} \beta^{-2/3}.
\eeq
Here, we have used the relation $Hx^{1/2}=2/(3t_\Lambda\sqrt{\Omega_M})$,
where all quantities (including the matter density parameter $\Omega_M$) 
are evaluated at the present time, and for our numerical estimate we have 
taken $z_{rec}=1000$, $t_{cb}=3\cdot 10^{10}yr$, $H_0 = 100\ h\ km\ s^{-1}\
Mpc^{-1}$, with $h=.7$, and $\Omega_M=.3$. For $\beta\sim 1$, as
suggested by the distribution (\ref{short}), we have that the most
likely values of $\sigma_{rec}$ are of the order of $10^{-3}$. This
is close to the observationally suggested values
$\sigma_{rec}=(3.3-2.4) 10^{-3}$ \cite{Martel}.

\begin{figure}[t]
\centering
\hspace*{-4mm}
\leavevmode\epsfysize=10 cm \epsfbox{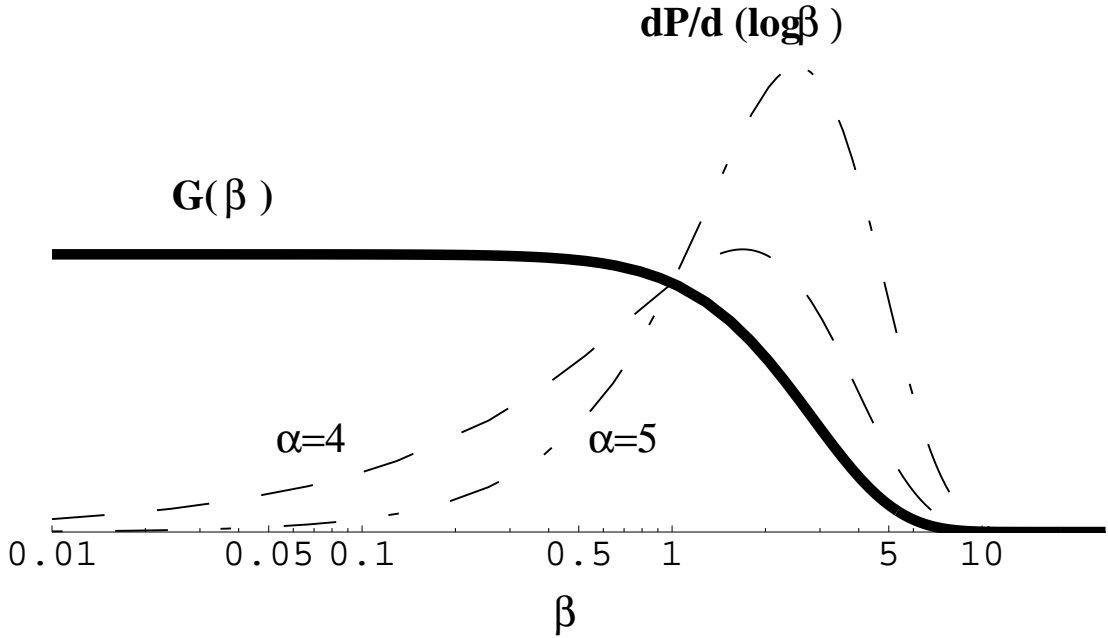}\\[3mm]
\caption[fig4]{\label{fig4} 
The probability density per unit logarithmic interval
of $\beta\propto \sigma_{rec}^{-3/2}$, Eq. (\ref{short}), for $\alpha=4$ and
$5$ (thind solid lines). The function $G(\beta)$ is represented by a 
thick solid line. }
\end{figure}

Anthropic bounds on the density contrast have been recently
discussed by Tegmark and Rees \cite{Tegmark}.  Instead of
$\sigma_{rec}$, they used the amplitude of the density fluctuations at
horizon crossing, $Q$; the relation between the two is roughly $Q\sim
10^{-2}\sigma_{rec}$.  They imposed a lower bound on $Q$ by requiring
that galaxies form prior to the ``cooling boundary'',
$t_\sigma\lesssim t_{cb}$.  This gives $Q\gtrsim 10^{-6}$.  To obtain
an upper bound, it has been argued \cite{RESCEU,Tegmark} that for
large values of $Q$ galaxies would be too dense and frequent stellar
encounters would disrupt planetary orbits.  To estimate the rate of
encounters, the relative stellar velocity was taken to be the virial
velocity $v_{vir}\sim 200 ~km~s^{-1}$, resulting in a bound $Q\lesssim
10^{-4}$.  However, Silk \cite{Silk} has pointed out that the local
velocity dispersion of stars in our Galaxy is an order of magnitude
smaller than $v_{vir}$.  This gives $Q\lesssim 10^{-3}$, which is a
rather weak constraint.  This issue does not arise in the approach
we take in the present paper, since in our case large values of $Q$
are suppressed by the {\it a priori} distribution ${\cal
P}_*(\sigma_{rec})$. 

\subsection{The time coincidence}

Finally, we should check that the introduction of a cooling boundary 
does not spoil the coincidence $t_G \sim t_\Lambda$. In fact, this seems
rather clear from Fig. 2. Introducing the cooling boundary basically 
amounts to disregarding the probability density above a certain horizontal 
line $t_G=t_{cb}$. The  probability
distribution for $t_G/t_\Lambda$ below the horizontal line is
somewhat different from that 
in the whole plane, but clearly it still peaks 
around $t_G \sim t_\Lambda$. To quantify this effect, we have integrated 
(\ref{prob2}) with respect to $x_{rec}$ over the range
$\sinh(x_{rec}^{1/2} t_{cb}/t_{rec})>x^{1/2}$. 
The resulting distribution for $x$ is proportional to the integrand
in the right hand side of (\ref{gdistr}).
For $\beta=2$, this probability density is shown in Fig. 1 
(curve $b$). The peak is only slightly shifted towards  
smaller values of $t_G/t_\Lambda$.

Cooling failure is not the only mechanism that can in principle inhibit
the number of civilizations at low $\sigma_{rec}$.  It is possible,
for example, that the stellar initial mass function (IMF) depends on the
protogalactic density $\rho_{vir}$, so that the number of carbon
forming stars drops rapidly towards very low values of 
$\rho_{vir}$.  If the a
priori distribution ${\cal P}_*(\sigma_{rec})$ is a decreasing
function of $\sigma_{rec}$, this can result in a peaked distribution
$d{\cal P}/d \ln t_\Lambda$.  Quite similarly, if the number of
relevant stars grows towards smaller $\rho_{vir}$, a peaked
distribution is obtained for an increasing function ${\cal
P}_*(\sigma_{rec})$.  Our present understanding of star formation is
insufficient to determine the dependence of the IMF
on $\rho_{vir}$, but once it is understood, the probability
distribution for $t_\Lambda$ can be calculated as outlined above
\cite{Footnew}.

\section{Conclusions}

In this paper we suggested a possible explanation for the
near-coincidence 
of the three cosmological timescales: the time of galaxy formation $t_G$, the
time when the cosmological constant starts to dominate the energy density
of the universe $t_\Lambda$, and the present age of the universe
$t_0$. Since this 
coincidence involves specifically the time of our existence as observers, it
lends itself most naturally to the consideration of anthropic selection
effects. 

We considered a model in which the cosmological constant is a
random variable 
with a flat a priori probability distribution. We showed that a
typical galaxy 
in this model forms at a time $t_G\sim t_\Lambda$. 
We further demonstrated that a typical
civilization should determine the value of the cosmological constant at 
$t_0\sim t_G$. Thus we should not be surprised to find ourselves
discussing the cosmic time coincidence.

     We also considered a model in which both the cosmological constant
$\Lambda$ and the density contrast $\sigma_{rec}$ are random variables.
The galaxy formation in this case is spread over a much wider
time interval, and we had to account for the fact that the cooling
of protogalactic clouds collapsing at very late times is too slow to
allow for efficient fragmentation and star formation.  We therefore
disregarded all galaxies formed after the ``cooling boundary'' time
$t_{cb}$.  We assumed that the {\it a priori} distribution for
$\sigma_{rec}$ is a decresing power law and found that, for a
sufficiently steep power, a typical observer detects $\sigma_{rec}\sim
10^{-3}-10^{-4}$, close to the values inferred from observations.
Such observers are likely to find themselves living at $t_0\sim t_\Lambda$
in a galaxy formed at $t_G\sim t_\Lambda$ in a region of the universe
where $t_\Lambda\sim t_{cb}$, also close to the observationally
suggested value.


Our model with variable $\Lambda$ and $\sigma_{rec}$ can be
developed further in several directions. Instead of taking a flat
distribution for $\rho_\Lambda$ and a power-law distribution
for $\sigma_{rec}$, one could use the methods of
Refs. \cite{AV98,VVW} to calculate the {\it a priori} distributions for
these variables in the framework of some inflationary model.  One could
also use a more refined model of structure formation and improve on
our treatment of cooling failure, replacing the sharp cutoff at
$t=t_{cb}$ with a more realistic model.  We believe, however, that
even in the present, simplified form our model indicates that an anthropic
selection for $\Lambda$ and $\sigma_{rec}$ is a viable possibility.

     Finally, we should note that the coincidence in the timescales requires
an explanation even in models involving a quintessence component
\cite{Quintessence}.  In models
of quintessence the universe at late times is dominated by a scalar field
$\phi$, slowly evolving down its potential $V(\phi)$. It has been argued (by
Zlatev {\it et al.} \cite{Zlatev}) that such models do not suffer from
the cosmic time
coincidence problem, because the time $t_\phi$ of $\phi$-domination is not
sensitive to the initial conditions. This time, however, does depend on the
details of the potential $V(\phi)$, and observers should be surprised to find
themselves living at the epoch when quintessence is about to dominate. More
satisfactory would be a model in which the potential depends on two fields,
say $\phi$ and $\chi$, with $\chi$ slowly varying in space, making the time of 
$\phi$-domination position dependent. Such models are not difficult to
construct 
in the context of inflationary cosmology. One could then apply the principle
of mediocrity to determine the most likely value of $t_\phi$.

\section*{Acknowlegements}

We are grateful to Ken Olum for his comments on the manuscript.
J.G. acknowledges support from CIRIT grant 1998BEAI400244.
M.L. acknowleges support from 
NASA Grant NAG5-6857.  A.V. was supported in part by the 
National Science Foundation.

\section*{Appendix: The probability distribution for $\Lambda$}

In this Appendix we briefly discuss the probability distribution for the
cosmological constant, giving a simplified version of the 
calculation presented in \cite{Martel}.

In a universe where the cosmological constant is non-vanishing, 
a primordial overdensity will eventually collapse provided that
its value at the time of recombination exceeds a certain critical value
$\delta^{rec}_{c}$. In the spherical collapse model this is estimated as 
$\delta^{rec}_{c}=1.13\  x_{rec}^{1/3}$ (see e.g. \cite{masha}). 
Hence, the fraction of matter that eventually clusters in galaxies 
can be roughly approximated as \cite{PS,masha}:
\begin{equation}
\nu(x_{rec})\approx 
{\rm erfc}\left({\delta^{rec}_c\over \sqrt{2}\sigma_{rec}(M_g)}\right)
\approx 
{\rm erfc}\left({.80\ x_{rec}^{1/3}\over\sigma_{rec}(M_g)}\right).
\label{nu}
\end{equation}
Here, erfc is the complementary error function and
$\sigma_{rec}(M_g)$ is the dispersion in the density contrast at the
time of recombination on the relevant galactic mass scale $M_g \sim 10^{12}
M_{\odot}$. 
The logarithmic distribution $d{\cal P}/d\ln\ x_{rec}=x_{rec}\nu(x_{rec})$ 
is plotted in Fig. 5.
It has a rather broad peak which spans two orders of magnitude in $x_{rec}$,
with a maximum at 
\begin{equation}
x^{peak}_{rec}\approx 2.45\ \sigma_{rec}^3.
\label{xpeak}
\end{equation}
In accordance with the principle of mediocrity, we should expect to 
measure a value of the cosmological constant within this broad peak of
the distribution. And indeed, this may actually be the case. 
The distribution (\ref{nu}) is characterized by the parameter $\sigma_{rec}$.
As noted by Martel et al. \cite{Martel}, this parameter can be inferred 
from observations of the cosmic microwave background anisotropies, although
its value depends on the assumed value of the cosmological constant today.  
For instance, assuming that the present cosmological constant is
$\Omega_{\Lambda,0}=.8$, and the relevant galactic co-moving scale is in 
the range $R= (1-2) Mpc$, Martel et al. found 
$\sigma_{rec} = (2.3-1.7)\ 10^{-3}$. In this estimate, they also assumed a
scale invariant spectrum of density perturbations, a value
of $70 km\ s^{-1}\ Mpc^{-1}$ for the present Hubble rate,  and they defined
recombination to be at redshift $z_{rec}\approx 1000$ (this definition is 
conventional, since the probability distribution for the cosmological 
constant does not depend on the choice of reference time). Thus, taking
into account that $x$ scales like $(1+z)^{-3}$ in equation (\ref{xpeak}),
one finds that the peak of the distribution for the cosmological constant
today is at $x_0^{peak} \approx 29.8 - 12$. The value corresponding
to the assumed $\Omega_{\Lambda,0}=.8$ is $x_0 = 4$, certainly within
the broad peak of the distribution and not far from its maximum. 
If instead we assume that the measured value is $\Omega_{\Lambda,0}=.7$, 
which corresponds to $x_0= 2.33$, the new inferred values for $\sigma_{rec}$
correspond to the peak value $x_0^{peak}\approx (88 - 34)$. In this case,
the measured value would be at the outskirts of the broad peak, where the
logarithmic probability density is about an order of magnitude smaller 
than at the peak, but still significant. Thus, even though there may be
uncertainties in the inferred value of $\sigma_{rec}$ on the relevant scales,
it seems fair to say that any observed value of $\Omega_{\Lambda,0} 
\gtrsim .7$ is in good agreement with the principle of mediocrity.

\begin{figure}[t]
\centering
\hspace*{-4mm}
\leavevmode\epsfysize=10 cm \epsfbox{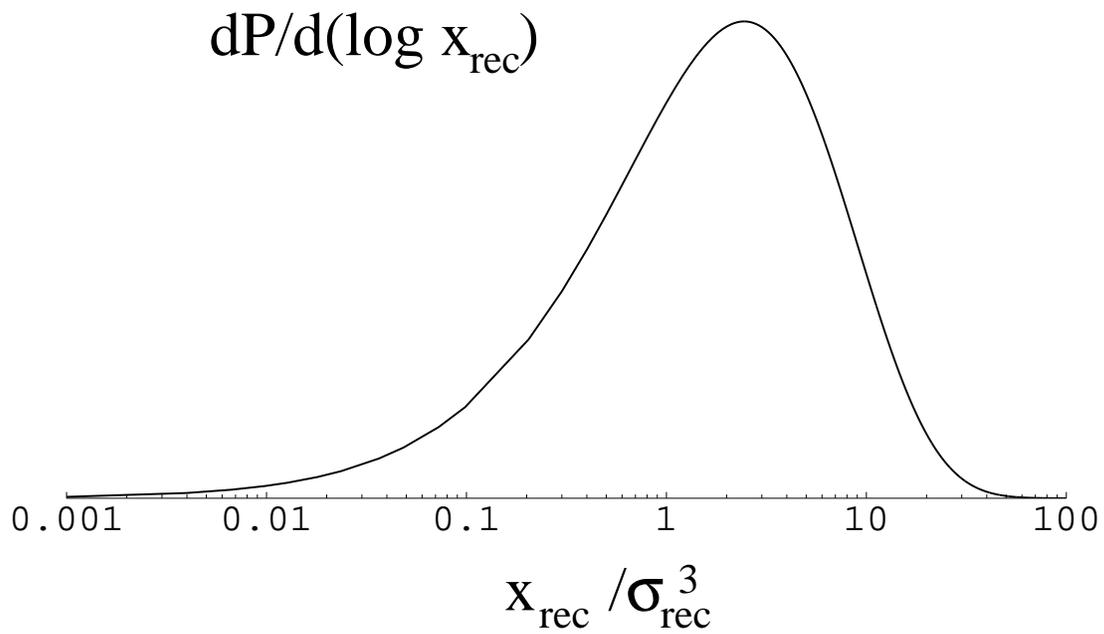}\\[3mm]
\caption[fig5]{\label{fig5} The probability density per unit logarithmic
interval of $x_{rec} \sigma_{rec}^{-3}$.
 }
\end{figure}








\begin{thebibliography}{99}
\bibitem{Foot1}
A similar argument for the coincidence of the galaxy formation time
$t_G$ with the curvature domination time $t_\Omega$ in models of open
inflation was given in Ref.\cite{GTV}.
\bibitem{Footnew}
There exist some indications that the IMF may depend on density.  For
example, it has been suggested that the IMF is biased towards high
masses in very active regions like starbursts \cite{Rieke}, and
towards low masses in regions of extreme inactivity, like the field
population \cite{Massey}.  These suggestions are not very certain
because the observations are influenced by selection effects and by
evolution.
\bibitem{Foot4}
Actually, the fraction
of baryonic matter $X_b$ is also relevant for cooling.
Following \cite{Tegmark} we shall take $X_b \sim 0.1$.
\bibitem{Foot5}
This upper bound on $t$ is determined by line cooling in Helium.
For $M_g \approx 10^{12} M_{\odot}$ there is also a narrow interval
near $t \approx 3 \cdot 10^{11} {\rm yr}$, 
where cooling is again efficient 
due to hydrogen line cooling. However, the interval is very narrow
and we shall disregard the galaxies which may form during this short 
late period.
\end{thebibliography}
\end{document}